\def\OPTIONConf{1}%
\def\OPTIONArxiv{1}%

\ifnum\OPTIONConf=2
  \documentclass[runningheads,a4paper]{llncs}
   \usepackage[breaklinks]{hyperref}
   \usepackage{breakurl}
   \usepackage{jdunfield}
   \usepackage{amssymb}

   \usepackage{amsmath}
   \usepackage{amsthm}
   \usepackage{thmtools}
   \usepackage{thm-restate}
   \usepackage{etoolbox}
\usepackage[T1]{fontenc}
\usepackage[authoryear]{natbib}
\else
\if\OPTIONArxiv=0
    \documentclass[format=acmsmall, amsthm=true,
      review=true,
      screen=true,
      printccs=false,
      anonymous=true,
      printacmref=false]%
      {acmart}
\else
    \documentclass[format=acmsmall, amsthm=true,
      review=false,
      screen=true,
      printccs=false,
      anonymous=false, nonacm=true,
      printacmref=false]%
      {acmart}
\fi
    \def~{\nobreakspace{}}
    \setcopyright{none}
\fi
\makeatletter
\renewcommand{\bibsection}{%
     \section*{\refname\@mkboth{\MakeUppercase{\refname}}{\MakeUppercase{\refname}}}%
}
\makeatother
\bibpunct{(}{)}{;}{a}{}{,}

\ifnum\OPTIONConf=1
 \usepackage{thmtools}
 \usepackage{thm-restate}
\fi

\declaretheoremstyle[
  bodyfont=\sl
]{mytheoremstyle}

\theoremstyle{mytheoremstyle}

\ifnum\OPTIONConf=1
  \newtheorem{theorem}{Theorem}
  \newtheorem{lemma}{Lemma}

  \theoremstyle{definition}

  \theoremstyle{remark}
\fi

\usepackage{fontenc}

\usepackage{etoolbox}
\usepackage{tcolorbox}

\usepackage{srcltx}

\usepackage{llproof}
\usepackage{rulelinks}

\usepackage{relsize}

\usepackage{latexsym}
\ifnum\OPTIONConf=0
  \usepackage{stmaryrd}
\fi

\usepackage{enumerate}

\ifnum\OPTIONConf=1
  \setcitestyle{nosort}    %
\else
\fi

\usepackage{pgf,tikz}
\usetikzlibrary{shapes,arrows,matrix,fit,backgrounds,calc,decorations,decorations.pathmorphing,decorations.markings}

\def\MathparLineskip{\lineskiplimit=0.9em\lineskip=0.75em plus 0.2em}

\usepackage{centernot}

\usepackage{semantic}   %

\usepackage{braket}     %
\usepackage{mathpartir} %
\usepackage{rotating} %
\usepackage{proof}
\usepackage{pdflscape}
\usepackage{fancyvrb} %
\usepackage{mathtools} %

\let\MathRightArrow\Rightarrow %
\usepackage{marvosym} %
\def\Rightarrow{\MathRightArrow}

\ifnum\OPTIONConf=2

\else
  \theoremstyle{remark}

\fi

\mathlig{|-->}{\longmapsto}

\newcommand{\chkcolor}{dDkBlue}
\newcommand{\syncolor}{dDkRed}
\newcommand{\isyncolor}{dDkGreen}
\newcommand{\chk}{\mathrel{\mathcolor{\chkcolor}{\Leftarrow}}}
\newcommand{\uncoloredsyn}{{\Rightarrow}}
\newcommand{\syn}{\mathrel{\mathcolor{\syncolor}{\uncoloredsyn}}}
\newcommand{\RRightarrow}{\mathrel{\mathrlap{\Rightarrow}\mkern2mu\Rightarrow}}
\newcommand{\isyn}{\mathrel{\mathcolor{\isyncolor}{\RRightarrow}}}
\mathlig{=>}{\syn}
\mathlig{=>>}{\isyn}
\mathlig{<=}{\chk}

\newcommand{\synchksym}{%
        \text{\raisebox{0.8ex}[0pt]{%
                        \ensuremath{\printsavewidth{{\syn}}\unskipsavedwidth\hspace*{-0.2ex}}}%
                \raisebox{-0.8ex}[0pt]{%
                        \ensuremath{{\chk}}}%
}}
\newcommand{\synchk}{\mathrel{\synchksym}}

\mathlig{<=>}{\synchk}

\makeatletter
\newsavebox{\@brx}
\newcommand{\llangle}[1][]{\savebox{\@brx}{\(\m@th{#1\langle}\)}%
  \mathopen{\copy\@brx\kern-0.5\wd\@brx\usebox{\@brx}}}
\newcommand{\rrangle}[1][]{\savebox{\@brx}{\(\m@th{#1\rangle}\)}%
  \mathclose{\copy\@brx\kern-0.5\wd\@brx\usebox{\@brx}}}
\makeatother

\newcommand{\Bool}{\tyname{bool}}

\newcommand{\Int}{\tyname{int}}

\newcommand{\Rat}{\tyname{rat}}
\newcommand{\rat}[2]{{#1} / {#2}}

\newcommand{\Nat}{\tyname{nat}}

\ifdefined\OPTIONTalk
\colorlet{sectcolor}{dBlue}
\colorlet{uncolor}{dRed}
\colorlet{diacolor}{dPurple}
\else
\colorlet{sectcolor}{black}
\colorlet{uncolor}{black}
\colorlet{diacolor}{black}
\fi
\newcommand{\topty}{\top}
\newcommand{\untysym}{\vee}
\newcommand{\unty}{\mathbin{\untysym}}
\newcommand{\secttysym}{\wedge}
\newcommand{\sectty}{\mathbin{\secttysym}}

\newcommand{\diasymsym}{\Diamond}
\newcommand{\diasym}{\mathcolor{diacolor}{\diasymsym}}
\newcommand{\dia}{\mathbin{\diasym}}
\newcommand{\hunty}{\dia}

\newcommand{\ite}[3]{\keyword{if}\;{#1}\;\keyword{then}\;{#2}\;\keyword{else}\;{#3}}

\newcommand{\RuleHead}[1]{\text{\raisebox{1em}[0pt]{\ensuremath{\mathsz{\ifnum\OPTIONConf=1 14pt\else 18pt \fi}{#1}}}}~~~~~}

\newcommand{\unitty}{\tyname{unit}}

\newcommand{\recsym}{\mu}

\newcommand{\sub}{\subtype}

\newrulecommand{DCons}{\rulename{DirCons}}  %

\newcommand{\subPf}[3]{\Pf{#1}{\subtype}{#2}{#3}}

\newcommand{\deepsubPf}[3] {\Pf{#1}{:<}{#2}{#3}}

\newcommand{\stsPf}[3] {\mkpf{\sts}{#1}{#2}{#3}}

\newcommand{\xproj}[1]{\keyword{proj}_{#1}}
\newcommand{\proj}[1]{\xproj{#1}\,}

\newcommand{\pair}[2]{\Pair{#1}{#2}}

\newcommand{\stepsym}{{\mapsto}}
\newcommand{\step}{\mathrel{\stepsym}}

\mathlig{++}{\mathbin{\mathbin{\texttt{\bf \upshape +}}}}

\newcommand{\rulename}[1]{\text{\normalfont\textsf{#1}}}

\newcommand{\subrulename}[1]{\rulename{$<:${#1}}}
\newrulecommand{SubRefl}{\subrulename{Refl}}
\newrulecommand{SubInt}{\subrulename{$\Int$}}
\newrulecommand{SubBool}{\subrulename{$\Bool$}}
\newrulecommand{SubFun}{\subrulename{${->}$}}

\newrulecommand{SubProd}{\subrulename{$*$}}
\newrulecommand{SubRec}{\subrulename{$\recsym$}}
\newrulecommand{SubRecL}{\subrulename{$\recsym$L}}
\newrulecommand{SubRecR}{\subrulename{$\recsym$R}}
\newrulecommandONE{SubSectL}{\subrulename{$\sectty$L$_{#1}$}}
\newrulecommand{SubSectR}{\subrulename{$\sectty$R}}

\newcommand{\bcdsubrulename}[1]{\rulename{${<:}_{\text{bcd}}${#1}}}
\newrulecommand{bcdsubrefl}{\bcdsubrulename{refl}}

\newrulecommand{bcdsubTopR}{\bcdsubrulename{TopRight}}
\newrulecommand{bcdsubTopL}{\bcdsubrulename{TopLeft}}
\newrulecommand{bcdsubSectR}{\bcdsubrulename{SectR}}
\newrulecommandONE{bcdsubSectL}{\bcdsubrulename{SectL{#1}}}

\newrulecommand{bcdsubdist}{\bcdsubrulename{dist}}

\newrulecommand{bcdsubtrans}{\bcdsubrulename{trans}}

\newrulecommand{bcdsubSectCong}{\bcdsubrulename{SectCong}}
\newrulecommand{bcdsubArr}{\bcdsubrulename{$->$}}

\newcommand{\bcdrulename}[1]{(\rulename{#1})}
\newcommand{\bcdintroname}[1]{\bcdrulename{{#1}I}}
\newcommand{\bcdelimname}[1]{\bcdrulename{{#1}E}}

\newrulecommand{bcdvar}{\bcdrulename{var}}

\newrulecommand{bcdArrIntro}{\bcdintroname{$->$}}
\newrulecommand{bcdArrElim}{\bcdelimname{$->$}}

\newrulecommand{bcdSectIntro}{\bcdintroname{$\sectty$}}
\newrulecommandONE{bcdSectElim}{\bcdrulename{$\sectty$E{#1}}}

\newrulecommand{bcdTopIntro}{\bcdintroname{$\topty$}}

\newrulecommand{bcdsub}{\bcdrulename{$<:$}}

\newrulecommand{bcdbetaeta}{\bcdrulename{$\beta\eta$}}

\newcommand{\bstarsym}{{{\entails}^{*}_{\beta\eta}}}
\newcommand{\bstar}{\mathbin{\bstarsym}}

\newcommand{\ered}{\rightarrow_{\eta}}
\newcommand{\ereds}{\twoheadrightarrow_{\eta}}
\newcommand{\bereds}{\twoheadrightarrow_{\beta\eta}}

\newcommand{\eredsPf}[3] {\mkpf{\ereds}{#1}{#2}{#3}}

\newcommand{\bstarPf}[3]{\Pf{#1}{\bstar}{#2}{#3}}

\newcommand{\TpBetaEta}{(\ensuremath{\beta\eta})\xspace}

\newcommand{\eredname}[1]{eta-{#1}}
\newrulecommand{eredArr}{\eredname{$->$}}
\newrulecommand{eredProd}{\eredname{$*$}}

\newcommand{\eredsname}[1]{etas-{#1}}
\newrulecommand{eredsOne}{\eredsname{1}}
\newrulecommand{eredsZero}{\eredsname{0}}
\newrulecommand{eredsTrans}{\eredsname{trans}}
\newrulecommand{eredsLam}{\eredsname{lam}}
\newrulecommand{eredsApp}{\eredsname{app}}
\newrulecommand{eredsOp}{\eredsname{op}}
\newrulecommand{eredsIte}{\eredsname{ite}}
\newrulecommand{eredsAnno}{\eredsname{anno}}
\newrulecommand{eredsPair}{\eredsname{pair}}
\newrulecommand{eredsProj}{\eredsname{proj}}

\newcommand{\srcchktypecolor}[1]{{#1}}
\newcommand{\srcsyntypecolor}[1]{{#1}}

\newcommand{\srcchkrulename}[1]{\rulename{\srcchktypecolor{Chk#1}}}
\newcommand{\srcsynrulename}[1]{\rulename{\srcsyntypecolor{Syn#1}}}

\newrulecommand{ChkUnitIntro}{\srcchkrulename{UnitIntro}}
\newrulecommand{ChkIntIntro}{\srcchkrulename{IntIntro}}
\newrulecommand{SynVar}{\srcsynrulename{Var}}

\newrulecommand{ChkSub}{\srcchkrulename{CSub}}

\newrulecommand{SynNonprincipal}{\srcsynrulename{Nonprincipal}}  

\newrulecommand{SynAnno}{\srcsynrulename{Anno}}  
\newrulecommand{ChkFunIntro}{\srcchkrulename{$->$Intro}}
\newrulecommand{SynFunElim}{\srcsynrulename{$->$Elim}}
\newrulecommandONE{ChkSumIntro}{\srcchkrulename{$+$Intro$_{#1}$}}
\newrulecommand{ChkSumElim}{\srcchkrulename{$+$Elim}}
\newrulecommand{ChkSumElimOne}{\srcchkrulename{$+$ElimOne}}
\newrulecommand{ChkProdIntro}{\srcchkrulename{$*$Intro}}
\newrulecommandONE{SynProdElim}{\srcchkrulename{$*$Elim$_{#1}$}}

\newrulecommandONE{ChkSubsumIntro}{\srcchkrulename{$\plussym_{#1}$Intro}}
\newrulecommandONE{ChkSubsumElim}{\srcchkrulename{$\plussym_{#1}$Elim}}

\newrulecommand{ChkRecIntro}{\srcchkrulename{$\recsym$Intro}}
\newrulecommand{SynRecElim}{\srcsynrulename{$\recsym$Elim}}
\newrulecommand{ChkSectIntro}{\srcchkrulename{$\sectty$Intro}}
\newrulecommandONE{SynSectElim}{\srcsynrulename{$\sectty$Elim$_{#1}$}}
\newrulecommand{ChkUnionElim}{\srcchkrulename{UnionElim}}
\newrulecommandONE{ChkUnionIntro}{\srcchkrulename{$\unty$Intro$_{#1}$}}

\newrulecommand{ChkPossIntro}{\srcchkrulename{PossIntro}}
\newrulecommand{ChkPossIntroVac}{\srcchkrulename{PossIntroVac}}
\newrulecommand{SynPossElim}{\srcsynrulename{PossElim}}

\newcommand{\srctypecolor}[1]{{#1}}
\newcommand{\srctyperulename}[1]{\rulename{\srctypecolor{S#1}}}
\newcommand{\srcintrorulename}[1]{\srctyperulename{{#1}Intro}}
\newcommand{\srcelimrulename}[1]{\srctyperulename{{#1}Elim}}

\newrulecommand{SUnitIntro}{\srcintrorulename{$\unitty$}}
\newrulecommand{SVar}{\srctyperulename{Var}}

\newrulecommand{SSub}{\srctyperulename{Sub}}

\newrulecommand{SAnno}{\srctyperulename{Anno}}  
\newrulecommand{SFunIntro}{\srcintrorulename{$->$}}
\newrulecommand{SFunElim}{\srcelimrulename{$->$}}
\newrulecommand{SSumIntro}{\srcintrorulename{Sum}}
\newrulecommand{SSumElimOne}{\srcelimrulename{Sum}\rulename{\srctypecolor{1}}}
\newrulecommand{SSumElimTwo}{\srcelimrulename{Sum}\rulename{\srctypecolor{2}}}

\newrulecommand{SUnionIntro}{\srcintrorulename{$\unty$}}
\newrulecommand{SUnionElim}{\srcelimrulename{$\unty$}}
\newrulecommand{SSectIntro}{\srcintrorulename{$\sectty$}}
\newrulecommand{SSectElim}{\srcelimrulename{$\sectty$}}
\newrulecommand{SHUnionIntro}{\srcintrorulename{$\hunty$}}
\newrulecommand{SHUnionElim}{\srcelimrulename{$\hunty$}}

\newcommand{\srcsttypecolor}[1]{\textcolor{black}{#1}}
\newcommand{\srcsttyperulename}[1]{\rulename{\srcsttypecolor{St#1}}}
\newcommand{\srcstintrorulename}[1]{\srcsttyperulename{{#1}Intro}}
\newcommand{\srcstelimrulename}[1]{\srcsttyperulename{{#1}Elim}}

\newrulecommand{SSUnitIntro}{\srcstintrorulename{$\unitty$}}
\newrulecommand{SSVar}{\srcsttyperulename{Var}}

\newrulecommand{SSSub}{\srcsttyperulename{Sub}}

\newrulecommand{SSAnno}{\srcsttyperulename{Anno}}  
\newrulecommand{SSFunIntro}{\srcstintrorulename{$->$}}
\newrulecommand{SSFunElim}{\srcstelimrulename{$->$}}
\newrulecommand{SSSumIntro}{\srcstintrorulename{$+$}}
\newrulecommand{SSSumElim}{\srcstelimrulename{$+$}}
\newrulecommand{SSProdIntro}{\srcstintrorulename{$*$}}
\newrulecommand{SSProdElim}{\srcstelimrulename{$*$}}

\newrulecommand{SSSectIntro}{\srcstintrorulename{$\sectty$}}
\newrulecommand{SSSectElim}{\srcstelimrulename{$\sectty$}}
\newrulecommand{SSUnionIntro}{\srcstintrorulename{$</$}}
\newrulecommand{SSUnionElim}{\srcstelimrulename{$</$}}
\newrulecommand{SSRecIntro}{\srcstintrorulename{$\recsym$}}
\newrulecommand{SSRecElim}{\srcstelimrulename{$\recsym$}}

\newcommand{\srcdyntypecolor}[1]{\textcolor{black}{#1}}
\newcommand{\srcdyntyperulename}[1]{\rulename{\srcdyntypecolor{D#1}}}
\newcommand{\srcdynintrorulename}[1]{\srcdyntyperulename{{#1}Intro}}
\newcommand{\srcdynelimrulename}[1]{\srcdyntyperulename{{#1}Elim}}

\newrulecommand{SDUnitIntro}{\srcdynintrorulename{$\unitty$}}
\newrulecommand{SDVar}{\srcdyntyperulename{Var}} 

\newrulecommand{SDSub}{\srcdyntyperulename{Sub}} 

\newrulecommand{SDAnno}{\srcdyntyperulename{Anno}} 
\newrulecommand{SDFunIntro}{\srcdynintrorulename{$->$}}
\newrulecommand{SDFunElim}{\srcdynelimrulename{$->$}}
\newrulecommand{SDSumIntro}{\srcdynintrorulename{$+?$}}
\newrulecommand{SDSumElimOne}{\srcdynelimrulename{$+?$}\rulename{\srcdyntypecolor{1}}}
\newrulecommand{SDSumElimTwo}{\srcdynelimrulename{$+?$}\rulename{\srcdyntypecolor{2}}}

\newcommand{\targettypecolor}[1]{\textcolor{dBlue}{#1}}
\newcommand{\targettyperulename}[1]{\rulename{\targettypecolor{T#1}}}
\newcommand{\targetintrorulename}[1]{\targettyperulename{{#1}Intro}}
\newcommand{\targetelimrulename}[1]{\targettyperulename{{#1}Elim}}

\newrulecommand{TUnitIntro}{\targetintrorulename{Unit}}
\newrulecommand{TIntIntro}{\targetintrorulename{Int}}
\newrulecommand{TVar}{\targettyperulename{Var}}
\newrulecommand{TFixVar}{\targettyperulename{FixVar}}
\newrulecommand{TFix}{\targettyperulename{Fix}}
\newrulecommand{TAbort}{\targettyperulename{Abort}}

\newrulecommand{TSome}{\targettyperulename{Some}}
\newrulecommand{TNone}{\targettyperulename{None}}
\newrulecommand{TOptionElim}{\targetelimrulename{Option}}
\newrulecommand{Toptionfail}{\targettyperulename{optionfail}}

\newrulecommand{TSub}{\targettyperulename{Sub}} 

\newrulecommand{TFunIntro}{\targetintrorulename{$->$}}
\newrulecommand{TFunElim}{\targetelimrulename{$->$}}
\newrulecommandONE{TSumIntro}{\targettyperulename{$+$Intro$_{#1}$}}
\newrulecommand{TSumElim}{\targetelimrulename{$+$}}
\newrulecommand{TProdIntro}{\targetintrorulename{$*$}}
\newrulecommandONE{TProdElim}{\targettyperulename{$*$Elim$_{#1}$}}

\newrulecommand{TRecIntro}{\targetintrorulename{$\recsym$}}
\newrulecommand{TRecElim}{\targetelimrulename{$\recsym$}}

\newrulecommandONE{TProSumIntroOne}{\targettyperulename{$\ps$Intro$_{#1}$}}
\newrulecommand{TProSumIntro}{\targetintrorulename{$\ps$}}
\newrulecommandONE{TProSumElim}{\targettyperulename{$\ps$Elim$_{#1}$}}
\newrulecommand{TPar}{\targettyperulename{Par}}

\newcommand{\transtypecolor}[1]{\textcolor{black}{#1}}
\newcommand{\transtyperulename}[1]{\rulename{\transtypecolor{ST#1}}}
\newcommand{\transintrorulename}[1]{\transtyperulename{{#1}Intro}}
\newcommand{\transelimrulename}[1]{\transtyperulename{{#1}Elim}}

\newrulecommand{STCSub}{\transtyperulename{CSub}}

\newrulecommand{STUnitIntro}{\transintrorulename{Unit}}
\newrulecommand{STIntIntro}{\transintrorulename{Int}}
\newrulecommand{STVar}{\transtyperulename{Var}}

\newrulecommand{STAnno}{\transtyperulename{Anno}}  
\newrulecommand{STFunIntro}{\transintrorulename{$->$}}
\newrulecommand{STFunElim}{\transelimrulename{$->$}}
\newrulecommandONE{STSumIntro}{\transtyperulename{$+$Intro$_{#1}$}}
\newrulecommandONE{STSubsumIntro}{\transtyperulename{$\plussym_{#1}$Intro}}
\newrulecommand{STSumElim}{\transelimrulename{$+$}}
\newrulecommandONE{STSubsumElim}{\transelimrulename{$\plussym_{#1}$}}
\newrulecommand{STProdIntro}{\transintrorulename{$*$}}
\newrulecommandONE{STProdElim}{\transtyperulename{$*$Elim$_{#1}$}}

\newrulecommand{STRecIntro}{\transintrorulename{$\recsym$}}
\newrulecommand{STRecElim}{\transelimrulename{$\recsym$}}

\newrulecommand{STSectIntro}{\transintrorulename{$\sectty$}}
\newrulecommandONE{STSectElim}{\transtyperulename{$\sectty$Elim$_{#1}$}}
\newrulecommandONE{STUnionIntro}{\transtyperulename{$</$Intro$_{#1}$}}
\newrulecommand{STUnionElim}{\transelimrulename{$</$}}

\newcommand\genrulename[1]{\rulename{{#1}}}
\newcommand\genintrorulename[1]{\genrulename{{#1}Intro}}
\newcommand\genelimrulename[1]{\genrulename{{#1}Elim}}

\newcommand\deeprulename[1]{\rulename{deep{#1}}}
\newcommand\shallowrulename[1]{\rulename{shallow{#1}}}
\newcommand\deepintrorulename[1]{\deeprulename{{#1}Intro}}
\newcommand\deepelimrulename[1]{\deeprulename{{#1}Elim}}

\newcommand\sts{|-_{\hspace*{-0.25ex}:<:}}

\newrulecommand{DeepVar}{\deeprulename{Var}}
\newrulecommand{DeepArrIntro}{\deepintrorulename{$->$}}
\newrulecommand{DeepArrElim}{\deepelimrulename{$->$}}
\newrulecommand{DeepBoolIntro}{\deepintrorulename{\Bool}}
\newrulecommand{DeepBoolElim}{\deepelimrulename{\Bool}}
\newrulecommand{DeepProdIntro}{\deepintrorulename{$*$}}
\newrulecommand{DeepProdElim}{\deepelimrulename{$*$}}

\newrulecommand{DeepSub}{\deeprulename{Sub}}
\newrulecommand{ShallowSub}{\shallowrulename{Sub}}

\newrulecommand{DeepAnno}{\deeprulename{Anno}}

\newrulecommand{DeepRat}{\deepelimrulename{\Rat}}
\newrulecommand{DeepIntIntro}{\deepintrorulename{\Int}}
\newrulecommand{DeepInt}{\deepelimrulename{\Int}}

\newrulecommand{GenVar}{\genrulename{Var}}
\newrulecommand{GenArrIntro}{\genintrorulename{$->$}}
\newrulecommand{GenArrElim}{\genelimrulename{$->$}}
\newrulecommand{GenBoolIntro}{\genintrorulename{\Bool}}
\newrulecommand{GenBoolElim}{\genelimrulename{\Bool}}
\newrulecommand{GenProdIntro}{\genintrorulename{$*$}}
\newrulecommand{GenProdElim}{\genelimrulename{$*$}}

\newrulecommand{GenDeepSub}{\genrulename{Sub}}

\newrulecommand{GenAnno}{\genrulename{Anno}}

\newrulecommand{GenRat}{\genelimrulename{\Rat}}
\newrulecommand{GenIntIntro}{\genintrorulename{\Int}}
\newrulecommand{GenInt}{\genelimrulename{\Int}}

\definecolor{deepcolor}{rgb}{0.3,0.3,0.9}

\ifnum\OPTIONConf=2
  \newcommand{\deepsubsym}{\textcolor{deepcolor}{\texttt{\upshape\selectfont<:}}}
\else
  \newcommand{\deepsubsym}{\textcolor{deepcolor}{\texttt{\upshape\selectfont<:\hspace*{-0.15ex}}}}
\fi
\newcommand{\deepsub}{\mathrel{\deepsubsym}}

\definecolor{shallowcolor}{rgb}{0.1,0.6,0.1}

\newcommand{\shallowsubsym}{\mathcolor{shallowcolor}{\leqslant}}
\newcommand{\shallowsub}{\mathrel{\shallowsubsym}}

\newcommand{\deepsubrulename}[1]{\rulename{DeepSub{#1}}}
\newcommand{\shallowsubrulename}[1]{\rulename{ShallowSub{#1}}}

\newrulecommand{DeepSubReflInt}{\deepsubrulename{ReflInt}}
\newrulecommand{DeepSubReflBool}{\deepsubrulename{ReflBool}}
\newrulecommand{DeepSubReflRat}{\deepsubrulename{ReflRat}}
\newrulecommand{DeepSubIntRat}{\deepsubrulename{IntRat}}
\newrulecommand{DeepSubArr}{\deepsubrulename{Arr}}
\newrulecommand{DeepSubProd}{\deepsubrulename{Prod}}

\newrulecommand{ShallowSubRefl}{\shallowsubrulename{Refl}}
\newrulecommand{ShallowSubIntRat}{\shallowsubrulename{IntRat}}

\mathlig{->}{\rightarrow}
\mathlig{**}{\times}

\mathlig{<:}{\sub}

\mathlig{:<}{\deepsub}
\mathlig{:<:}{\shallowsub}

\mathlig{<>}{\hunty}

\newcommand{\steprulename}[1]{\rulename{Step{#1}}}
\newrulecommand{StepContext}{\steprulename{Context}}
\newrulecommand{StepParL}{\steprulename{ParL}}
\newrulecommand{StepParR}{\steprulename{ParR}}
\newrulecommand{StepAbort}{\steprulename{Abort}}

\newcommand{\stepRrulename}[1]{\rulename{Reduce{#1}}}
\newrulecommand{ReduceBeta}{\stepRrulename{$\beta$}}
\newrulecommand{ReduceProj}{\stepRrulename{Proj}}
\newrulecommand{ReduceCase}{\stepRrulename{Case}}
\newrulecommand{ReduceSome}{\stepRrulename{Some}}
\newrulecommand{ReduceFail}{\stepRrulename{Fail}}

\newcommand{\anno}[2]{\texttt({#1}\;\texttt{:}\;{#2}\texttt)}

\newcommand{\lesscaptionspace}{\vspace*{-1.5ex}}

\ifdefined\OPTIONTalk
  
\else
  
\fi

\newcommand{\Btrue}{\datacon{True}}
\newcommand{\Bfalse}{\datacon{False}}

\newcommand{\byih}{By i.h.\xspace}

\newcommand{\bopsym}{\oplus}
\newcommand{\bop}{\mathbin{\bopsym}}

\ifnum\OPTIONConf=1
  \newcommand{\fixproofspacing}{\vspace*{-1ex}}
\else
  \newcommand{\fixproofspacing}{}
\fi

\def\OPTIONAnonymous{0}

\begin{document}

    \title%
      {Flattening subtyping by eta expansion}

\ifnum\OPTIONAnonymous=0
    \author{Jana Dunfield}

\ifnum\OPTIONConf=2
\institute{Queen's University at Kingston, Canada
\\
\email{jd169@queensu.ca}
}
\else
\orcid{0000-0002-3718-3395}
\affiliation{%
  \institution{Queen's University}
  \streetaddress{School of Computing, Goodwin Hall 557}
  \city{Kingston, ON}
  \postcode{K7L 3N6}
  \country{Canada}
}
\email{jd169@queensu.ca}
\fi
\fi

\ifnum\OPTIONConf=2
    \maketitle              %

\begin{abstract}
  To design type systems that use subtyping, we have to make tradeoffs.
  Deep subtyping is more expressive than shallow subtyping,
  because deep subtyping compares the entire structure of types.
  However, shallow subtyping is easier to reason about.
  By $\eta$-expanding source programs,
  we can get the effect of deep subtyping
  with less of its complexity.

  An early paper on filter models \citep{Barendregt83}
  examined two similar intersection type systems.
  The first included a subsumption rule that used
  a rich subtyping relation,
  including multiple rules for the top type
  and a distributivity rule.
  Their second type system dropped the subsumption rule,
  but added a rule that allowed a term to be
  $\eta$-expanded before typing it.
  This rule in their second type system
  compensated for the lack of subsumption:
  where their first type system used subtyping
  to manipulate intersections deep inside types,
  their second type system used introduction
  and elimination rules directly on the subterms
  created by $\eta$-expansion.
  Viewed as a computation, their proof of completeness for the second (shallow) system
  performs $\eta$-expansion.
  Thus, we can regard their proof as inventing the application of $\eta$-expansion
  to avoid deep subtyping.

  This paper serves as a tutorial on using $\eta$-expansion to obviate
  deep subtyping,
  puts the invention of the technique by \citet{Barendregt83}
  into context,
  gives a complete proof of the relevant lemma,
  and discusses how the technique can be used in
  type system design.
\end{abstract}

\else  %
\begin{abstract}
  To design type systems that use subtyping, we have to make tradeoffs.
  Deep subtyping is more expressive than shallow subtyping,
  because deep subtyping compares the entire structure of types.
  However, shallow subtyping is easier to reason about.
  By $\eta$-expanding source programs,
  we can get the effect of deep subtyping
  with less of its complexity.

  An early paper on filter models \citep{Barendregt83}
  examined two similar intersection type systems.
  The first included a subsumption rule that used
  a rich subtyping relation,
  including multiple rules for the top type
  and a distributivity rule.
  Their second type system dropped the subsumption rule,
  but added a rule that allowed a term to be
  $\eta$-expanded before typing it.
  This rule in their second type system
  compensated for the lack of subsumption:
  where their first type system used subtyping
  to manipulate intersections deep inside types,
  their second type system used introduction
  and elimination rules directly on the subterms
  created by $\eta$-expansion.
  Viewed as a computation, their proof of completeness for the second (shallow) system
  performs $\eta$-expansion.
  Thus, we can regard their proof as inventing the application of $\eta$-expansion
  to avoid deep subtyping.

  This paper serves as a tutorial on using $\eta$-expansion to obviate
  deep subtyping,
  puts the invention of the technique by \citet{Barendregt83}
  into context,
  gives a complete proof of the relevant lemma,
  and discusses how the technique can be used in
  type system design.
\end{abstract}

    \maketitle
\fi

\section{Introduction}
\label{sect:intro}

When we design a typed programming language, the status of the subtyping relation matters.
The full design space of subtyping is large.
An extreme position is to say that a type is a subtype of itself only,
in effect, having no subtyping at all;
another extreme position is to make subtyping as powerful as possible.
Within the setting of functional programming languages,
at the first extreme, we have the simply typed lambda calculus;
around the second extreme, we have semantic subtyping \citep{Frisch02}.%

Most type systems fall somewhere between these extremes,
even those that tend to be regarded as having no subtyping.
For example, taking polymorphism as a form of subtyping \citep{Odersky96},
the prefix polymorphism of Standard ML \citep{RevisedDefinitionOfStandardML}
is a form of subtyping.
We \emph{can} reasonably take polymorphism to be subtyping:
substitutability (presented by \citet{Liskov94} for object-oriented languages,
but also applicable in functional languages) says that if we expect a value of the supertype,
it is safe to use a value of the subtype:
if a function's domain is $\Int$,
we can safely pass a value of type $\Nat$, because every natural number is an integer.
If a function expects a function of type $\Int -> \Int$, it is safe to pass a function of type
$\forall \alpha.\,\alpha -> \alpha$.
But our use of terminology is not universal, now or (especially) in the past:
what we %
call subtyping has been called containment by \citep{Mitchell88,Remy05}.

A subtyping relation is \emph{shallow} if it examines only the heads of the types;
a subtyping relation is \emph{deep} if it examines the entire type.
Suppose we want the natural numbers to be
a subtype of the integers.
We can conclude $\Nat \subtype \Int$ just from the heads of the types
($\Nat$ is an atomic type, so its head is itself),
but to conclude
$(\Bool -> \Nat) \subtype (\Bool -> \Int)$
we need to go under the head connective $->$
and check that $\Nat \subtype \Int$.
So $\Nat \subtype \Int$ is a shallow subtyping,
while 
$(\Bool -> \Nat) \subtype (\Bool -> \Int)$
is a deep subtyping.

Choosing whether the subtyping relation \emph{visible to the programmer} is shallow or deep
is a language design decision that involves issues such as
programmer convenience, readability, and maintainability.
The choice also raises questions of elegance and simplicity of implementation:
deep subtyping is generally more complicated to implement.
It is also more complicated to reason about:
for example, deep subtyping needs more subtyping rules,
and thus more cases in proofs.

This paper is meant to serve as a tutorial on using
$\eta$-expansion
to elaborate a source language with deep subtyping
into a simpler language with only shallow subtyping,
in which we can do the rest of our reasoning (and compilation). 
A secondary purpose is to closely examine the origin of the technique:
the proof of a lemma in a paper by
\citet{Barendregt83},
who did not focus on the technique
and may not have noticed that they had invented it.
Their proof of the lemma showed only one case;
to make sure the result is valid, and in case any further insights lurk, we give a complete proof.

\begin{itemize}
\item \Sectionref{sec:system} presents a small language and type system with deep subtyping.

\item \Sectionref{sec:shallow} removes deep subtyping, leaving only shallow subtyping.

\item \Sectionref{sec:eta} defines $\eta$-expansion,
  and proves that programs using deep subtyping can be $\eta$-expanded into programs that use only shallow subtyping.

\item \Sectionref{sec:origins} summarizes \citet{Barendregt83} and explains where they used the technique.

\item \Sectionref{sec:applications} discusses potential uses of 
   the technique.
\end{itemize}

The technique we discuss is certainly not new, even if we discount its somewhat hidden origin;
for example, \citet{Pfenning08} proposed using it for datasort refinements.
However, this paper offers a fresh perspective on the technique by
discussing it in more detail,
and more closely examining the history of the technique.

\let\Hand\relax

\section{Example system}
\label{sec:system}

\subsection{Overview}

Our setting is the simply typed lambda calculus, extended with products, booleans,
and numeric types.
The latter provide (atomic) subtyping:
the type $\Int$ of integers is a subtype of the type $\Rat$ of rationals.
This gives us a ``numeric tower'' of types
that contains just enough to illustrate the technique we are presenting,
but keeps the rules relatively simple.
(We initially had a type $\Nat$ of natural numbers, instead of $\Rat$,
but $\Nat$ forces us to add a typing rule for literals as well as for arithmetic operators,
whereas $\Rat$ only needs new typing rules for arithmetic operators.)

\begin{figure}[bhtp]
  \centering

  \begin{grammar}
    Variables & $x, y, z$
    \\
    Integers & $n$ &\bnfas& $0 \bnfalt 1 \bnfalt -1 \bnfalt 2 \bnfalt -2 \bnfalt \cdots$
    \\
    Binary operators & $\bop$ &\bnfas& ${+} \bnfalt {-} \bnfalt {<} \bnfalt {/}$
    \\
    Product indexes & $k$ &\bnfas& $1 \bnfalt 2$
    \\[0.3ex]
    Types & $A, B$ &\bnfas&
          $
             \Int \bnfalt
             \Rat \bnfalt
             \Bool \bnfalt
             A -> B \bnfalt
             A * B 
          $
    \\[0.3ex]
    Expressions & $e$ &\bnfas&
          $x
           \bnfalt \lam{x} e \bnfalt e_1 e_2
           \bnfaltBRK  \anno{e}{A}
           \bnfaltBRK n \bnfalt e_1 \bop e_2
           \bnfaltBRK \Btrue \bnfalt \Bfalse \bnfalt \ite{e}{e_1}{e_2}
           \bnfaltBRK \pair{e_1}{e_2} \bnfalt \proj{k} e
           $
  \end{grammar}
  \lesscaptionspace
  
  \caption{Simply typed lambda calculus plus integers, rationals, booleans and products}
  \label{fig:stlc}
\end{figure}

\subsection{Grammar}
\Figureref{fig:stlc} gives the grammar of our variant of the STLC.
We write $x$, $y$ and $z$ for variables, and $n$ for integers.
The nonterminal $\oplus$ stands for any of the
binary operators: addition $+$, subtraction $-$,
comparison $<$, and division $/$ (which acts as the introduction form for the $\Rat$ type).
The nonterminal $k$ ranges over $1$ and $2$.

We write $A$ and $B$ for types:
integers $\Int$,
rationals $\Rat$,
booleans $\Bool$,
functions $A -> B$
and products $A * B$.

Expressions consist of
variables $x$, functions $\lam{x} e$,
function application $e_1 e_2$,
annotated expressions $\anno{e}{A}$,
integer literals $n$, binary operations $e_1 \oplus e_2$,
Boolean literals $\Btrue$ and $\Bfalse$,
conditional expressions $\ite{e}{e_1}{e_2}$,
pairs $\pair{e_1}{e_2}$,
and projections $\proj{k}{e}$.

Unlike lambdas with annotated bound variables ($\lam{x{:}A} e$),
the annotation form $\anno{e}{A}$ is general:
it can be used to annotate any expression form,
not only lambda.
Moreover, the general form works nicely with bidirectional typing.

\begin{figure}[thbp]
  
  \judgbox{\arrayenvbl{
      \Gamma |- e : A
      }}{
      Under typing assumptions $\Gamma$, expression $e$ has type $A$
    }
  \begin{mathpar}
    \Infer{\GenVar}
      {(x : A) \in \Gamma}
      {\Gamma |- x : A}
    ~
    \Infer{\GenArrIntro}
      {
        \Gamma, x : A |-  e : B
      }
      {\Gamma |- \lam{x} e : A -> B}
    ~
    \Infer{\GenArrElim}
      {
        \Gamma |- e_1 : A -> B
        \\
        \Gamma |- e_2 : A
      }
      {\Gamma |- e_1 e_2 : B}
  \end{mathpar}

  \begin{minipage}[t]{0.38\linewidth}
   ~\\[-16ex]
  \begin{mathpar}
    \Infer{\DeepSub}
        {
          \Gamma |- e : A
          \\
          A :< B
        }
        {\Gamma |- e : B}
  \end{mathpar}
  \end{minipage}
  \!\!\!\!
 \begin{minipage}[t]{0.61\linewidth}
  \begin{tcolorbox}[colback=green!1,
                  colframe=dDkGreen,
                  width=\linewidth,
                  arc=1mm, auto outer arc,
                 ]
     \vspace*{-1.3ex}
     \hspace*{-3.5ex} \judgbox{\arrayenvbl{
          \Gamma \sts e : A
          }}{
          Modification replacing
          ${:<}$ with ${:<:}$ %
        }
  \vspace{3ex}
  \hspace*{-2.5ex}Change $|-$ to $\sts$ throughout, and replace \DeepSub:\!\!\!\!
  \\[-1ex]
  \begin{mathpar}
    \Infer{\ShallowSub}
        {
          \Gamma \sts e : A
          \\
          A :<: B
        }
        {\Gamma \sts e : B}
  \vspace*{-1.2ex}
  \end{mathpar}
  \end{tcolorbox}
  \end{minipage}

\ifnum\OPTIONConf=2
  \vspace*{-3.2ex}
\else
  \vspace*{-1.4ex}
\fi

  \begin{mathpar}
    \Infer{\GenAnno}
        {
          \Gamma |- e : A
        }
        {
          \Gamma |- \anno{e}{A} : A
        }
   ~~
    \Infer{\GenProdIntro}
        {
          \Gamma |- e_1 : B_1
          \\
          \Gamma |- e_2 : B_2
        }
        {\Gamma |- \pair{e_1}{e_2} : B_1 * B_2}
    ~
    \Infer{\GenProdElim}
        {
          \Gamma |- e : A_1 * A_2
        }
        {
          \Gamma |- \proj{k}{e} : A_k
        }
    \\
    \Infer{\GenIntIntro}
      {}
      {\Gamma |- n : \Int}
    \and
    \Infer{\GenInt}
      {
         \Gamma |- e_1 : \Int
         \\
         \Gamma |- e_2 : \Int
       }
       {
         \Gamma |- e_1 + e_2 : \Int
         \\
         \Gamma |- e_1 - e_2 : \Int
         \\
         \Gamma |- e_1 < e_2 : \Bool
       }
    \\
    \Infer{\GenRat}
      {
         \Gamma |- e_1 : \Rat
         \\
         \Gamma |- e_2 : \Rat
       }
       {
         \Gamma |- e_1 + e_2 : \Rat
         ~~~
         \Gamma |- e_1 - e_2 : \Rat
         ~~~
         \Gamma |- e_1 < e_2 : \Bool
         ~~~
         \Gamma |- \rat{e_1}{e_2} : \Rat
       }
    \\
    \Infer{\GenBoolIntro}
      {b \in \{\Btrue, \Bfalse\} }
      {\Gamma |- b : \Bool}
    ~~~
    \Infer{\GenBoolElim}
        {
          \Gamma |- e : \Bool
          \\
          \arrayenvbl{
          \Gamma |- e_1 : B
          \\
          \Gamma |- e_2 : B
          }
        }
        {\Gamma |- \ite{e}{e_1}{e_2} : B}
\end{mathpar}

  \lesscaptionspace
    
  \caption{Rules for the system $|-$ using deep subtyping, and for $\sts$ using shallow subtyping}
  \label{fig:typing}
\end{figure}

\begin{figure}[thbp]
  \centering
  
  \judgbox{A :< B}{$A$ is a deep subtype of $B$}
  \vspace*{-1.0ex}
  \begin{mathpar}
    \Infer{\DeepSubReflInt}
       {}
       {\Int :< \Int}
    ~
    \Infer{\DeepSubReflBool}
       {}
       {\Bool :< \Bool}
    ~
    \Infer{\DeepSubReflRat}
       {}
       {\Rat :< \Rat}
   \vspace*{-1.2ex}
    \\
    \Infer{\DeepSubIntRat}
       {}
       {\Int :< \Rat}
    \\
    \Infer{\DeepSubArr}
       {
         B_1 :< A_1
         \\
         A_2 :< B_2
       }
       {(A_1 -> A_2) :< (B_1 -> B_2)}
    \and
    \Infer{\DeepSubProd}
       {
         A_1 :< B_1
         \\
         A_2 :< B_2
       }
       {(A_1 * A_2) :< (B_1 * B_2)}
  \end{mathpar} \\ \lesscaptionspace

  \caption{Subtyping rules (deep; compare \Figureref{fig:subtyping-shallow})}
  \label{fig:subtyping-deep}
\end{figure}

\begin{figure}[thbp]
  \centering
  
  \judgbox{A :<: B}{$A$ is a shallow subtype of $B$}
  \vspace*{-1.2ex}
  \begin{mathpar}
    \Infer{\ShallowSubRefl}
       {}
       {A :<: A}
    \and
    \Infer{\ShallowSubIntRat}
       {}
       {\Int :<: \Rat}
  \end{mathpar} \\ \lesscaptionspace
  
  \caption{Subtyping rules (shallow; compare Figure \ref{fig:subtyping-deep})}
  \label{fig:subtyping-shallow}
\end{figure}

\subsection{Typing rules} 
Rules \GenVar, \GenArrIntro, and \GenArrElim comprise the simply typed lambda calculus.
Rules \GenProdIntro and \GenProdElim extend the STLC with products.

Rule \GenIntIntro gives literals the type $\Int$,
and rule \GenInt types addition and subtraction.
We could write separate rules for $+$ and $-$,
but since they would have the same premises,
we compress them into one rule.
We similarly compress into \GenRat the rules for four operators
($+$, $-$, $/$ and $<$) 
on \emph{rational numbers} (instead of integers):
all require both arguments to have type $\Rat$.
Finally, we include rules \GenBoolIntro and \GenBoolElim,
so we can do something with the result of the $<$ operator.

\paragraph{Bidirectional version:}
Due to the subsumption rule, the typing rules in Figure \ref{fig:typing} do not immediately yield an algorithm.
As reassurance that this example type system could be implemented,
we give a bidirectional version of it (in the appendix).
To use Lemma \ref{lem:excore} and Theorem \ref{thm:eta}
with the bidirectional system,
we need only take the bidirectional derivation and replace all occurrences of
the synthesis ($=>$) and checking ($<=$) symbols with colons.

\subsection{Deep subtyping}
\label{sec:deep}

We write $A :< B$ for deep subtyping (Figure \ref{fig:subtyping-deep}).

The rule \DeepSubIntRat is important because it makes this subtyping relation
more than just reflexivity.

The rules \DeepSubArr and \DeepSubProd allow this relation (unlike the shallow one
described below) to look within subformulas of the given types:
\DeepSubArr is the usual such rule, contravariant in the domain
($\Int :< \Rat$, and a function that expects $\Rat$ may safely be used
in a context expecting $\Int$, so we want $(\Rat -> B) :< (\Int -> B)$) and covariant in the codomain.
\DeepSubProd is covariant (since, for example, $\Rat * \Int$ should be 
a subtype of $\Rat * \Rat$).

The remaining rules capture reflexivity of atomic types
(\DeepSubReflInt and similar rules for $\Bool$ and $\Rat$).

\subsection{Shallow subtyping}
\label{sec:shallow}

We write $A :<: B$ for \emph{shallow} subtyping (Figure \ref{fig:subtyping-shallow}).
If we exclude all the non-atomic subtyping rules,
only the rule concluding $\Int :<: \Rat$ remains (renamed \ShallowSubIntRat).
With only that one rule, reflexivity is not admissible,
so we add a rule \ShallowSubRefl that concludes $A :<: A$.
In addition to reflexivity being an expected property of subtyping,
the bidirectional version of the type system (Appendix \ref{sec:bidir})
uses reflexivity in the subsumption rule to switch between synthesis and checking.

\subsection{Eta expansion}
\label{sec:eta}

  \begin{figure}[!b]
    \centering

    \judgbox{e \ered e'}{Term $e$ locally eta-reduces to $e'$ \ifnum\OPTIONConf=2 \\ \fi
    (read from right to left: $e'$ locally eta-expands to $e$)}

    \begin{mathpar}
      \Infer{\eredArr}
         {
           x \notin \FV{e}
         }
         {\lam{x} (e\;x) \ered e}
      \and
      \Infer{\eredProd}
         {}
         {\pair{\proj{1}{e}}{\proj{2}{e}} \ered e}
    \end{mathpar}

    \judgbox{e \ereds e'}{Term $e$ eta-reduces (anywhere, zero or more times) to $e'$ \\
    (equivalently: $e'$ eta-expands to $e$)}

    \begin{mathpar}
      \Infer{\eredsOne}
         {e \ered e'}
         {e \ereds e'}
      \and
      \Infer{\eredsZero}
         {}
         {e \ereds e}
      \and
      \Infer{\eredsTrans}
         {
           e_1 \ereds e_2
           \\
           e_2 \ereds e_3
         }
         {e_1 \ereds e_3}
      \and
      \Infer{\eredsLam}
         {
           e \ereds e'
         }
         {\lam{x} e \ereds \lam{x} e'}
      \and
      \Infer{\eredsApp}
         {
           e_1 \ereds e_1'
           \\
           e_2 \ereds e_2'
         }
         {e_1\,e_2 \ereds e_1'\,e_2'}
      \and
      \Infer{\eredsOp}
         {
           e_1 \ereds e_1'
           \\
           e_2 \ereds e_2'
         }
         {(e_1 \bop e_2) \ereds (e_1' \bop e_2')}
      ~~
      \Infer{\eredsIte}
         {
           e \ereds e'
           \\
           e_1 \ereds e_1'
           \\
           e_2 \ereds e_2'
         }
         {\ite{e}{e_1}{e_2} \ereds \ite{e'}{e_1'}{e_2'}}
      \and
      \Infer{\eredsAnno}
         {
           e \ereds e'
         }
         {
           \anno{e}{A}
           \ereds
           \anno{e'}{A}
         }
      \and
      \Infer{\eredsPair}
         {
           e_1 \ereds e_1'
           \\
           e_2 \ereds e_2'
         }
         {
           \pair{e_1}{e_2}
           \ereds
           \pair{e_1'}{e_2'}
         }
      \and
      \Infer{\eredsProj}
         {
           e \ereds e'
         }
         {
           \proj{k}{e}
           \ereds
           \proj{k}{e'}
         }
    \end{mathpar}
    
    \caption{Eta reduction (if reading right-to-left: eta expansion)}
    \label{fig:ex-ereds}
  \end{figure}

We write $e \ered e'$ to mean that the whole expression $e$
eta-reduces (once) to $e'$; equivalently, $e'$ eta-expands (once) to $e$.
There are two rules: \eredArr for functions, and \eredProd for products
(Figure \ref{fig:ex-ereds}).
The rule for functions requires that $x$ be not free in $e$:
when eta-expanding, we choose a fresh $x$.

The $\ered$ judgment is used only within the $\ereds$ judgment:
$e \ereds e'$ says that $e$ eta-reduces anywhere, any number of times.
The definition of $e \ereds e'$ is the reflexive-transitive-congruence closure of $\ered$.

\subsection{From deep to shallow}

Theorem \ref{thm:eta} says that
the shallow subtyping system (after eta-expanding)
is complete with respect to the deep subtyping system.
Its proof is straightforward, once we have Lemma \ref{lem:excore},
which says that the deep subsumption rule is admissible in the shallow system \emph{if}
we can eta-expand.
The plain admissibility of subsumption would be
``if $A :< B$ and $\Gamma \sts e : A$ then $\Gamma \sts e : B$'',
which does not hold since $\sts$ has only shallow subsumption;
we need to allow the eta-expansion of $e$ before assigning it type $B$.

\smallskip

\begin{lemma}[Shallow expansion] \label{lem:excore}
  If $A :< B$ and
    $\Gamma \sts e : A$
    then there exists $e'$ such that $e' \ereds e$
    and
    $\Gamma \sts e' : B$.
\end{lemma}
\fixproofspacing
\begin{proof}
  By induction on the derivation of $A :< B$.

  \begin{itemize}
  \item 
  If the derivation is concluded by a reflexivity rule
  (\DeepSubReflInt,
  \DeepSubReflBool, or
  \DeepSubReflRat), then $A = B$.
  Let $e' = e$.  Then $\Gamma \sts e : A$ is actually $\Gamma \sts e' : B$.
  
  \smallskip

\item
  If the derivation is concluded by \DeepSubIntRat, let $e' = e$.
  \\ By \ShallowSubIntRat, $\Int :<: \Rat$. By \ShallowSub, $\Gamma \sts e' : \Rat$.

  \smallskip

\item
  If the derivation is concluded by \DeepSubArr:

  Let $x$ be a fresh variable.

  \begin{llproof}
    \eqPf{A}{A_1 -> A_2}  {By inversion}
    \eqPf{B}{B_1 -> B_2}  {\ditto}
    \deepsubPf{B_1}{A_1}  {\ditto}
    \deepsubPf{A_2}{B_2}  {\ditto}
    \stsPf{\Gamma}{e : A_1 -> A_2}  {Given}
    \stsPf{\Gamma, x : B_1}{e : A_1 -> A_2} {By weakening}
    \decolumnizePf
    \stsPf{\Gamma, x : B_1}{x : B_1} {By \GenVar}
    \stsPf{\Gamma, x : B_1}{e'_x : A_1 \AND e'_x \ereds x} {By induction hypothesis}
    \stsPf{\Gamma, x : B_1}{e\;e'_x : A_2} {By \GenArrElim}
    \stsPf{\Gamma, x : B_1}{e_1 : B_2 \AND e_1 \ereds e\,e'_x} {\byih}
    \stsPf{\Gamma}{\lam{x} e_1 : B_1 -> B_2} {By \GenArrIntro}
    \eredsPf{e'_x} {x}  {Above}
    \eredsPf{e\;e'_x} {e\;x}  {By \eredsZero, \eredsApp}
    \eredsPf{e_1} {e\;e'_x}  {Above}
    \eredsPf{e_1} {e\;x}  {By \eredsTrans}
    \eredsPf{\lam{x} e_1} {\lam{x} (e\;x)}  {By \eredsLam}
    \eredsPf{\lam{x} (e\;x)} {e}  {By \eredArr and \eredsOne}
    \eredsPf{\lam{x} e_1} {e}  {By \eredsTrans}
    \LetPf{e'} {\lam{x} e_1}
  \end{llproof}

\item
  If the derivation is concluded by \DeepSubProd:

\begin{llproof}
    \eqPf{A}{A_1 * A_2}  {By inversion}
    \eqPf{B}{B_1 * B_2}  {\ditto}
    \deepsubPf{A_1}{B_1}  {\ditto}
    \deepsubPf{A_2}{B_2}  {\ditto}
    \stsPf{\Gamma}{e : A_1 * A_2}  {Given}
    \stsPf{\Gamma}{\proj{1}{e} : A_1}  {By \GenProdElim}
    \stsPf{\Gamma}{e'_1 : B_1
      \AND e'_1 \ereds \proj{1}{e}}  {\byih}
    \stsPf{\Gamma}{\proj{2}{e} : A_2}  {By \GenProdElim}
    \stsPf{\Gamma}{e'_2 : B_2
      \AND e'_2 \ereds \proj{2}{e}}  {\byih}
    \stsPf{\Gamma}{\pair{e'_1}{e'_2} : B_1 * B_2} {By \GenProdIntro}
    \proofsep
    \eredsPf{\pair{e'_1}{e'_2}} {\pair{\proj{1}{e}}{\proj{2}{e}}}  {By \eredsPair}
    \eredsPf{} {e}  {By \eredProd and \eredsOne}
    \eredsPf{\pair{e'_1}{e'_2}} {e}  {By \eredsTrans}
    \LetPf{e'} {\pair{e'_1}{e'_2}}
  \end{llproof}
  \qedhere
  \end{itemize}
\end{proof}

\begin{theorem}
\label{thm:eta}
~
If $\Gamma |- e : A$
then
there exists $e'$
such that $e' \ereds e$
and $\Gamma \sts e' : A$.
\end{theorem}
\fixproofspacing
\begin{proof}
  By induction on the derivation of $\Gamma |- e : A$.

  \begin{itemize}
  \item Case \GenVar: apply \GenVar.
  \item Case \GenArrElim:

    The induction hypothesis gives $e_1'$ and $e_2'$ such that
    $e_1' \ereds e_1$ and $\Gamma \sts e_1' : A_2 -> A$
    and $e_2' \ereds e_2$ and $\Gamma \sts e_2' : A_2$.
    
    By \GenArrElim, $\Gamma \sts e_1' e_2' : A -> B$.

    By rule \eredsApp, $e_1' e_2' \ereds e_1 e_2$.

  \item Cases \GenAnno,
 \GenProdIntro, \GenBoolElim,
 \GenProdElim, \GenIntIntro, \GenInt, \GenRat, \GenBoolIntro:
    Similar to the \GenArrElim case: apply the induction hypothesis to each premise,
    apply the corresponding rule of the $\sts$ system,
    and use the congruence of $\ereds$.
  \item Case \GenArrIntro: \\
     By inversion, $e = (\lam{x} e_0)$ and $A = A_1 -> A_2$ and $\Gamma, x{:}A_1 |- e_0' : A_2$.
    By the induction hypothesis, we have $e_0'$ such that
    $e_0' \ereds e_0$ and $\Gamma, x{:}A_1 \sts e_0' : A_2$.
    By \GenArrIntro, $\Gamma \sts (\lam{x} e_0') : A_1 -> A_2$.

  \item Case \DeepSub:

    \begin{llproof}
      \ePf{\Gamma}{e : A_1}  {Subderivation}
      \deepsubPf{A_1}{A}  {Subderivation}
      \stsPf{\Gamma}{e_1 : A_1}  {\byih}
      \eredsPf{e_1}{e}  {\ditto}
\Hand      \stsPf{\Gamma}{e' : A}  {By Lemma \ref{lem:excore} (Shallow expansion)}
      \eredsPf{e'}{e_1}  {\ditto}
\Hand      \eredsPf{e'}{e}  {By \eredsTrans}
    \end{llproof} \qedhere
  \end{itemize}
\end{proof}

\section{Origins}
\label{sec:origins}

In this section, we examine some of the type systems and key results from \citet{Barendregt83},
which we refer to as \emph{BCD}
for Barendregt, Coppo, and Dezani-Ciancaglini.
Their paper seems best known as a primarily, if not purely, theoretical exploration of filter models
using intersection types.

Our main concern is their Lemma 4.2,
but to provide context, we begin by summarizing their paper (\Secref{sec:bcd-summary}).
Section \ref{sec:bcd-extended} presents BCD's extended type assignment
($\lambda$-calculus extended with intersection and a universal, or ``top'', type).
Section \ref{sec:bcd-lemma-proof} gives a full proof of BCD's Lemma 4.2.

\let\Hand\relax

\subsection{Summary of BCD}
\label{sec:bcd-summary}

The central equivalence considered by \citet{Barendregt83} 
is that $M$ has type $\sigma$ in Curry type assignment---where terms do
not contain types, and the typing rules are just the usual two
$->$ rules and a $\beta$-equality rule---if and only if $M$ has type $\sigma$ in a $\lambda$-model.  We will call the latter \emph{semantic typing};
a term has type $\sigma$ in a $\lambda$-model
if it always behaves like (reduces to) something of type $\sigma$.
Since the lemma we focus on does not actually involve $\lambda$-models
or semantic typing, we do not repeat BCD's Definition 1.2 or the explanations given by \citet[pp.\ 57--58]{Hindley97}.

This central equivalence consists of

\begin{enumerate}[(1)]
\item \emph{soundness}: if $M$ has type $\sigma$ in Curry type assignment ($|-_{C}$),
  then $M$ has type $\sigma$ in a $\lambda$-model;
\item \emph{completeness}: if $M$ has type $\sigma$ in a $\lambda$-model,
  then there exists a derivation giving $M$ the type $\sigma$ in Curry type assignment.
\end{enumerate}

Soundness was already known: BCD cite the dissertation of \citet{Ben-YellesThesis}.
\citet[p.\ 59]{Hindley97} states
``4B8 Soundness'' with a one-word proof (``Straightforward''),
but offers a specific citation of \citeauthor{Ben-YellesThesis}' Theorem 4.17, which
appears, indeed, straightforward\footnote{%
The proof of Theorem 4.17 uses Ben-Yelles' Lemmas 4.14* and 4.16*,
which apply the logic of the $->$-introduction and $->$-elimination rules (respectively) within the $\lambda$-model.
(The asterisks are in the original, following a convention defined within it.)
}.
However, 4.17 is for a system without a $\beta$ typing rule,
so to actually get the result for the Curry type assignment system in BCD---%
which does have that rule---we must combine 4.17 \citep[p.\ 70]{Ben-YellesThesis}
with 4B4 (i) \citep[p.\ 58]{Hindley97}.

The main result of BCD is the completeness of Curry type assignment, %
which they prove as follows:

\begin{enumerate}
\item Define an \emph{extended type assignment} system (BCD Def.\ 2.5(i)) with intersection types $\sect$ and a universal (top) type $\omega$.  This system includes introduction and elimination rules for intersection, an introduction rule for $\omega$, drops the $\beta$-equality rule (admissible per BCD Corollary 3.8), and adds a subsumption rule.

\item Show completeness of extended type assignment with respect to semantic typing (BCD 3.10).

\item Delete the subsumption rule and add a rule \bcdbetaeta (BCD Definition 4.1(ii)), which
  says that if $M$ has type $\sigma$ and $M$ is $\beta\eta$-reducible
  to $N$, then $N$ has type $\sigma$.  (They also allow the typing context, or \emph{basis}, to be ``large'': it can include typing assumptions about terms that are not variables; see BCD Definition 4.1(i).
But this does not seem to be needed to prove Lemma 4.2, which is our focus.)

\item Show (BCD Lemma 4.2) that this system, which replaces subsumption with \bcdbetaeta,
  is complete with respect to extended type assignment.

\item Show (BCD Corollary 4.10) that Curry type assignment is complete
  with respect to extended type assignment:
  given an extended type assignment derivation,
  BCD Lemma 4.2 gives a derivation in the modified system,
  which can be normalized; by the subformula property,
  that derivation cannot use intersection and $\omega$
  (those features being the \emph{extended} part of extended type assignment),
  so it is a Curry type assignment derivation.
\end{enumerate}

\subsection{BCD's extended type assignment}
\label{sec:bcd-extended}

\citet{Barendregt83} $\eta$-convert within the proof of their Lemma 4.2,
which states that
if $M : \sigma$ in a system with a general subsumption rule (for a particular subtyping relation),
then $M : \sigma$ in a system that omits subsumption but includes a \TpBetaEta rule:
\[
   \Infer{$(\beta\eta)$}
       {
         \Gamma \bstar M : \sigma
         \\
         M \bereds N
      }
      {
        \Gamma \bstar N : \sigma
      }
\]
Their subtyping relation \citep[Def.\ 2.3(i)]{Barendregt83} is shown
(in Gentzen's rule notation) in \Figureref{fig:bcd-subtyping}.
Their ``extended type assignment theory''
is shown in \Figureref{fig:bcd-typing}.
We make several changes to notation:

\begin{figure}[t]
  \centering

  \vspace*{-0.7ex}
  \begin{mathpar}
    \Infer{\bcdsubrefl}
        {}
        {\tau <: \tau}
    \and
    \Infer{\bcdsubTopR}
        {}
        {\tau <: \top}
    \and
    \Infer{\bcdsubTopL}
        {}
        {\top <: (\top -> \top)}
    \vspace*{-0.7ex}
    \\
    \Infer{\bcdsubSectR}
        {}
        {\tau <: \tau \sectty \tau}
    \and
    \Infer{\bcdsubSectL{1}}
        {}
        {\sigma \sectty \tau <: \sigma}
    \and
    \Infer{\NoLinkbcdsubSectL{2}}
        {}
        {\sigma \sectty \tau <: \tau}
    \vspace*{-1.3ex}
    \\
    \Infer{\bcdsubdist}
        {}
        {(\sigma -> \tau_1) \sectty (\sigma -> \tau_2) <: \sigma -> (\tau_1 \sectty \tau_2)}
    \\
    \Infer{\bcdsubtrans}
        {
          \sigma_1 <: \sigma_2
          \\
          \sigma_2 <: \sigma_3
        }
        {\sigma_1 <: \sigma_3}
    \and
    \Infer{\bcdsubSectCong}
        {
          \sigma <: \sigma'
          \\
          \tau <: \tau'
        }
        {(\sigma \sectty \tau) <: (\sigma' \sectty \tau')}
    \and
    \Infer{\bcdsubArr}
         {
           \tau_1 <: \sigma_1
           \\
           \sigma_2 <: \tau_2
         }
         {
           (\sigma_1 -> \sigma_2) <: (\tau_1 -> \tau_2)
         }
  \end{mathpar}

  \lesscaptionspace
  
  \caption{BCD subtyping \citep[Def.\ 2.3(i)]{Barendregt83} in Gentzen's rule notation}
  \label{fig:bcd-subtyping}
\end{figure}

\begin{figure}[t]
  \centering

   \judgbox{\Gamma |- M : \tau}{Extended type assignment
   incl.\ \NoLinkbcdsub~\citep[Def.\ 2.5]{Barendregt83}}

  \begin{mathpar}
    \Infer{\bcdvar}
        {(x : \tau) \in \Gamma}
        {\Gamma |- x : \tau}
    ~
    \Infer{\bcdArrIntro}
        {\Gamma, x:\sigma |- M : \tau}
        {\Gamma |- (\lam{x} M) : (\sigma -> \tau)}
    ~
    \Infer{\bcdArrElim}
        {
          \Gamma |- M : (\sigma -> \tau)
          \\
          \Gamma |- N : \sigma
        }
        {\Gamma |- M\,N : \tau}
    \\
    \Infer{\bcdSectIntro}
        {
          \Gamma |- M : \sigma
          \\
          \Gamma |- M : \tau
        }
        {
          \Gamma |- M : (\sigma \sectty \tau)
        }
    ~~~
    \Infer{\bcdSectElim{1}}
        {
          \Gamma |- M : (\sigma \sectty \tau)
        }
        {
          \Gamma |- M : \sigma
        }
    ~~~
    \Infer{\bcdSectElim{2}}
        {
          \Gamma |- M : (\sigma \sectty \tau)
        }
        {
          \Gamma |- M : \tau
        }
    \vspace*{-1.5ex}
    \\
    ~~~~~~~~\Infer{\bcdTopIntro}
         {}
         {\Gamma |- M : \topty}
    \hfill~
  \end{mathpar}

\ifnum\OPTIONConf=2
  \vspace*{-6.5ex}
\else
  \vspace*{-4.5ex}
\fi

  \begin{minipage}[t]{0.32\linewidth}
   ~\\[-17ex]
  \begin{mathpar}
    \Infer{\bcdsub}
         {
           \Gamma |- M : \sigma
           \\
           \sigma <: \tau
         }
         {\Gamma |- M : \tau}
  \end{mathpar}
  \end{minipage}
  \!\!
 \begin{minipage}[t]{0.66\linewidth}
  \begin{tcolorbox}[colback=purple!1,
                  colframe=dPurple,
                  width=\linewidth,
                  arc=1mm, auto outer arc,
                 ]
    \vspace*{-1ex}
\hspace*{-3ex}
          \judgbox{\Gamma \bstar M : \tau}{
       {Modification replacing \bcdsub with \bcdbetaeta}
        \\ \citep[Def.\ 4.1(ii)]{Barendregt83}}

      \vspace{2.7ex}

      Include all rules except \bcdsub, replacing $|-$ with $\bstarsym$
      \vspace*{-0.7ex}
      \begin{mathpar}
          \Infer{\bcdbetaeta}
             {
               \Gamma \bstar M : \sigma
               \\
               M \bereds N
             }
             {\Gamma \bstar N : \tau}
  \vspace*{-1.5ex}
      \end{mathpar}    
  \end{tcolorbox}
  \end{minipage}

  \caption{Adaptation of BCD's extended type assignment, and its modification
}
  \label{fig:bcd-typing}
\end{figure}

\begin{enumerate}
\item Following \citet{Curry58}, BCD wrote $\sigma M$ to mean that $M$ has type $\sigma$; we write $M : \sigma$.
\item 
BCD gave the rules in the style of natural deduction \citep{Gentzen35,Prawitz65}
with assumptions floating above their scope,
but used sequent notation $B |- \cdots$ to discuss derivations.
We state the rules using the judgment form of sequent calculus,
with typing assumptions $\Gamma$ written within the judgment
(so we add a rule \bcdvar).
\item
We use the judgment form $\Gamma |- \cdots$
for BCD's extended type assignment (which includes a subsumption rule with deep subtyping),
and the form $\Gamma \bstar \cdots$
for BCD's modified extended type assignment (which replaces the subsumption rule with the $(\beta\eta)$ rule).

\item BCD used $\cap$ for intersection
  and $\omega$ for top;
  we write $\sectty$ and $\topty$, respectively.

\end{enumerate}

\subsection{Proof of BCD Lemma 4.2}
\label{sec:bcd-lemma-proof}

We give a full proof of their Lemma 4.2.

\smallskip 

\begin{lemma}[{\citet[Lemma 4.2, p.\ 937]{Barendregt83}}]
If $\Gamma |- M : \sigma$ (including subsumption and excluding $\beta\eta$) then $\Gamma \bstar M : \sigma$ (excluding subsumption, including $\beta\eta$, and allowing $\Gamma$ to be \emph{large}).
\end{lemma}
\fixproofspacing
\fixproofspacing
\begin{proof}  By induction on the given derivation.
  
  Most rules of the $|-$-system are also rules of the $\bstar$-system.
  When such a shared rule concludes $\Gamma |- M : \sigma$,
  it suffices to apply the induction hypothesis\ to all premises, and then apply the same rule.

  The interesting case is when the subsumption rule concludes the derivation,
  which (according to \citeauthor{Barendregt83}) requires showing that $\sigma_1 \subtype \sigma_2$
  implies $M : \sigma_1 \bstar M : \sigma_2$.
  This is stated in the original proof as ``[t]he only thing to show'' and ``done by induction
  on the definition of $\leq$ using rule ($\beta\eta$)'' (with one case given).
  We state and prove this part separately, as Lemma \ref{lem:core} (Core of 4.2).  Then we have:

  \begin{llproof}
    \ePf{\Gamma}{M : \tau}  {Subderivation}
    \subPf{\tau}{\sigma}  {Subderivation}
    
    \bstarPf{\Gamma}{M : \sigma}   {By Lemma \ref{lem:core} (Core of 4.2)}
  \end{llproof}
 \end{proof}

The original Lemma 4.2 takes a derivation
of $B |- \sigma M$
and produces a derivation of $B \bstar %
\sigma M$.
We interpret the statement 
``\dots where we allow $B$ to be large'' (just above Lemma 4.2)
as saying that the input derivation
  $B |- \sigma M$ is in a modification of the system in the
early sections of \citet{Barendregt83} (since that system did not
allow large bases), perhaps to allow the proof to
create a large basis.
However, our proof never creates a large basis.
BCD write
(adapting notation): if $\sigma \subtype \tau$, then $M : \sigma \bstar %
M : \tau$.

Our version of this is: for all $B$, if $\sigma \subtype \tau$ and $B \bstar %
M : \sigma$ then $B \bstar %
M : \tau$.
When $B$ is not large, %
the two versions are essentially the same,
and since they use this result only within the proof of 4.2
(in fact, they ``inline'' its proof) where the basis may be non-empty,
there seems to be no important difference.

\begin{lemma}[Core of 4.2] \label{lem:core}
  If $\sigma \subtype \tau$ (\citet{Barendregt83} Def.\ 2.3(i))
  and $\Gamma \bstar M : \sigma$
  then $\Gamma \bstar M : \tau$.
\end{lemma}
\fixproofspacing
\begin{proof}
  By induction on the derivation of $\sigma \subtype \tau$.

  We consider cases of the rule concluding $\sigma \subtype \tau$.
  First, we consider the (only) case presented by \citet{Barendregt83}:

  \begin{itemize}
  \DerivationProofCase{\bcdsubArr}
         {
           \tau_1 <: \sigma_1
           \\
           \sigma_2 <: \tau_2
         }
         {
           \underbrace{\sigma_1 -> \sigma_2}_{\sigma}
           <:
           \underbrace{\tau_1 -> \tau_2}_{\tau}
         }

         \begin{llproof}
           \bstarPf{\Gamma} {M : \sigma_1 -> \sigma_2}   {Given}
           \bstarPf{\Gamma, x : \tau_1} {M : \sigma_1 -> \sigma_2}   {By weakening}
           \proofsep
           \bstarPf{\Gamma, x : \tau_1}{x : \tau_1} {By \bcdvar}
           \subPf{\tau_1}{\sigma_1}  {Subderivation}
           \bstarPf{\Gamma, x : \tau_1}{x : \sigma_1} {By induction hypothesis}
           \bstarPf{\Gamma, x : \tau_1}{M\;x : \sigma_2}  {By \bcdArrElim}
           \proofsep
           \subPf{\sigma_2}{\tau_2}  {Subderivation}
           \bstarPf{\Gamma, x : \tau_1}{M\;x : \tau_2}  {\byih}
           \proofsep
           \bstarPf{\Gamma}{(\Lam{x} M\;x) : \tau_1 -> \tau_2}  {By \bcdArrIntro}
           \Pf{\lam{x} (M\;x)}{\bereds}{M}  {$\eta$-reduction}
         \Hand  \bstarPf{\Gamma}{M : \tau_1 -> \tau_2}  {By \bcdbetaeta}
         \end{llproof}
  \end{itemize}

  We consider the remaining cases in the same order as the BCD subtyping definition \citep[Def.\ 2.3(i)]{Barendregt83},
  given (in rule form) in \Figureref{fig:bcd-subtyping}.

  \begin{itemize}
  \DerivationProofCase{\bcdsubrefl}
         {}
         {
           \underbrace{\tau}_{\sigma} <: \tau
         }

         Since $\tau = \sigma$, we already have our goal.

  \DerivationProofCase{\bcdsubTopR}
         { }
         {
           \sigma <: \underbrace{\top}_{\tau}
         }

         \begin{llproof}
         \Hand \bstarPf{\Gamma}{ M : \top}  {By \bcdTopIntro}
         \end{llproof}

  \DerivationProofCase{\bcdsubTopL}
         { }
         {
           \underbrace{\top}_{\sigma} <: \underbrace{\top -> \top}_{\tau}
         }

         \begin{llproof}
           \bstarPf{\Gamma, x : \top} {M\;x : \top}    {By \bcdTopIntro}
           \bstarPf{\Gamma} {\lam{x} (M\;x) : \top -> \top}    {By \bcdArrIntro}
           \Pf{\lam{x} (M\;x)}{\bereds}{M}  {$\eta$-reduction}
         \Hand  \bstarPf{\Gamma} {M : \top -> \top}    {By \bcdbetaeta}
         \end{llproof}

  \DerivationProofCase{\bcdsubSectR}
         { }
         {
           \sigma <: \underbrace{\sigma \sectty \sigma}_{\tau}
         }

         \begin{llproof}
           \bstarPf{\Gamma} {M : \sigma}  {Given}
           \Hand \bstarPf{\Gamma} {M : \sigma \sectty \sigma}    {By \bcdSectIntro}
         \end{llproof}

  \DerivationProofCase{\bcdsubSectL{1}}
         { }
         {
           \sigma_1 \sectty \sigma_2 <: \sigma_1
         }

         \begin{llproof}
           \bstarPf{\Gamma} {M : \sigma_1 \sectty \sigma_2}  {Given}
           \Hand \bstarPf{\Gamma} {M : \sigma_1}    {By \bcdSectElim{1}}
         \end{llproof}

  \DerivationProofCase{\bcdsubSectL{2}}
         { }
         {
           \sigma_1 \sectty \sigma_2 <: \sigma_2
         }

         Similar to the sect-left-1 case, but using \bcdSectElim{2}. %

  \DerivationProofCase{\bcdsubdist}
         { }
         {
           (\sigma_1 -> \tau_1)
           \sectty
           (\sigma_1 -> \tau_2)
           <:
           \sigma_1 -> (\tau_1 \sectty \tau_2)
         }

         \begin{llproof}
           \bstarPf{\Gamma} {M : (\sigma_1 -> \tau_1) \sectty (\sigma_1 -> \tau_2)}  {Given}
           \proofsep
           \bstarPf{\Gamma} {M : \sigma_1 -> \tau_1} {By \bcdSectElim{1}}
           \bstarPf{\Gamma, x : \sigma_1} {M : \sigma_1 -> \tau_1}    {By weakening}
           \bstarPf{\Gamma, x : \sigma_1} {x : \sigma_1}    {By \bcdvar}
           \bstarPf{\Gamma, x : \sigma_1} {M\;x : \tau_1}    {By \bcdArrElim}
           \decolumnizePf
           \bstarPf{\Gamma} {M : \sigma_1 -> \tau_2} {By \bcdSectElim{2}}
           \bstarPf{\Gamma, x : \sigma_1} {M : \sigma_1 -> \tau_2}    {By weakening}
           \bstarPf{\Gamma, x : \sigma_1} {x : \sigma_1}    {By \bcdvar}
           \bstarPf{\Gamma, x : \sigma_1} {M\;x : \tau_2}    {By \bcdArrElim}
           \proofsep
           \bstarPf{\Gamma, x : \sigma_1} {M\;x : \tau_1 \sectty \tau_2}    {By \bcdSectIntro}
           \bstarPf{\Gamma} {\lam{x} (M\;x) : \sigma_1 -> (\tau_1 \sectty \tau_2)}    {By \bcdArrIntro}
           \Pf{\lam{x} (M\;x)}{\bereds}{M}  {$\eta$-reduction}
           \Hand \bstarPf{\Gamma} {M : \sigma_1 -> (\tau_1 \sectty \tau_2)}    {By \bcdbetaeta}
         \end{llproof}

  \DerivationProofCase{\bcdsubtrans}
         {
           \sigma <: \tau'
           \\
           \tau' <: \tau
         }
         {
           \sigma <: \tau
         }

         \begin{llproof}
           \bstarPf{\Gamma} {M : \sigma}  {Given}
           \proofsep
           \subPf{\sigma}{\tau'}  {Subderivation}
           \bstarPf{\Gamma} {M : \tau'} {\byih}
           \proofsep
           \subPf{\tau'}{\tau}  {Subderivation}
         \Hand  \bstarPf{\Gamma} {M : \tau} {\byih}
         \end{llproof}

  \DerivationProofCase{\bcdsubSectCong}
         {
           \sigma_1 <: \tau_1
           \\
           \sigma_2 <: \tau_2
         }
         {
           \sigma_1 \sectty \sigma_2
           <:
           \tau_1 \sectty \tau_2
         }

         \begin{llproof}
           \bstarPf{\Gamma} {M : \sigma_1 \sectty \sigma_2}  {Given}
           \proofsep
           \bstarPf{\Gamma} {M : \sigma_1}  {By \bcdSectElim{1}}
           \subPf{\sigma_1}{\tau_1}  {Subderivation}
           \bstarPf{\Gamma} {M : \tau_1} {\byih}
           \proofsep
           \bstarPf{\Gamma} {M : \sigma_2}  {By \bcdSectElim{2}}
           \subPf{\sigma_2}{\tau_2}  {Subderivation}
           \bstarPf{\Gamma} {M : \tau_2} {\byih}
           \proofsep
         \Hand  \bstarPf{\Gamma} {M : \tau_1 \sectty \tau_2} {By \bcdSectIntro}
         \end{llproof}
\qedhere
  \end{itemize}
\end{proof}

\subsection{Discussion}

Two aspects of this proof seem worth noting.

First, while rule \TpBetaEta allows both $\beta$- and $\eta$-conversion,
the proof
uses only $\eta$-conversion.

Second, the premise of \bcdbetaeta allows $\eta$-reduction,
not expansion:
if $M$ has type $\sigma$
and $M$ $\beta\eta$-reduces to $N$,
then $N$ also has type $\sigma$.
But if we start from the conclusion and move to the premises---%
as type checkers do---%
then we start with a term $N$
and $\eta$-\emph{expand} it to get an $M$.

This suggests that
we can reduce the problem of type checking with deep subtyping
to the problem of type checking with shallow subtyping, %
provided that we can figure out where to $\eta$-expand.

\section{Applications}
\label{sec:applications}

The simplest way to use eta expansion to avoid most of the work of deep subtyping is
to allow deep subtyping in your type system for source programs,
and insert eta-expansions where deep subtyping is used.
Later compiler phases, then, only need shallow subtyping.

If we want to completely avoid implementing deep subtyping,
we must take another approach.
The definition of eta expansion may be as straightforward as that of beta reduction,
but knowing \emph{where} to eta-expand requires type information.
In many settings, this is taken care of already:
refinement type systems often assume that unrefined type inference has been done first \citep{DaviesThesis},
and dependent type systems may reconstruct approximate types first \citep{Pientka13}.
As long as the unrefined or approximate types tell us which terms have types
that need eta-expansion, we can eta-expand before beginning the ``real'' typing process
(by ``process'' we include type inference, type checking, and any combination of them).
Even when we do not have access to that information before we start the typing process,
we may get the information along the way:
for example, in a function application we know (and must check) that the function has $->$ type,
so if we see the same function variable used as an argument later, we will know to eta-expand it.
In short, knowing where to eta-expand requires type information
that may take some work to get,
except in Church-style settings with type information baked into the terms.

Our technique may be applicable to datasort refinement systems \citep{Davies00icfpIntersectionEffects,DaviesThesis,Dunfield03}:
the presence of intersection (and union) types makes subtyping (sometimes called subsorting) essential.
Since these systems restrict intersection to similar types---%
$R \sectty S$ is well formed only if $R$ and $S$ refine the same unrefined type $A$---%
the unrefined type information should tell us how to eta-expand enough.
In a small system of datasort refinements, \citet[\S 12]{Pfenning08} proposed avoiding
deep subtyping (subsorting) by eta expansion, but as far as we know
this has not been scaled up to a complete datasort refinement system.

On the other hand, in a system of \emph{unrestricted} intersection and union types \citep{Dunfield14},
completely avoiding deep subtyping seems difficult:
since the system is motivated by forming the intersections (and unions) of completely different types,
we have no unrefined type information and would not know where to eta-expand.

The existence of the technique of eta expansion
has been used to justify the omission of deep subtyping;
examples include a type system for incremental computation \citep{Chen14} and
a type system for GADTs \citep{Dunfield19}.
However, neither of these papers actually formulated eta-expansion
in their respective settings.

\citet{Mitchell88} gave two equivalent ways of adding subtyping to System F:
an eta rule (like BCD's $(\beta\eta)$ without the $\beta$ part),
and a subsumption (containment) rule.

Eta-conversion \citep[p.\ 92]{Curry58}
is perhaps almost as central as beta-conversion.
This paper is about eta-conversion \emph{for subtyping},
but we mention a few further applications here.
Expanding, instead of reducing, can be used to recover confluence in some type systems \citep{Jay95}; in the logical framework LF, terms are compared after converting them to
a beta-normal, eta-long canonical form \citep{Harper05}---%
who also note that defining eta-long form is not always straightforward.
While most uses of eta-expansion are in service of \emph{static} reasoning
(in type systems, or to reason about rewriting, to the extent those are actually different),
\citet{Rioux23} eta-expand in the \emph{dynamic} semantics.

%
%
%
%

\section{Conclusion}
\label{sec:conclusion}

We have presented a brief tutorial on a simple but useful application
of eta-expansion, describing how it works in an illustrative example type system.
We also put the first known (to us) development of the technique
into technical and historical context:
it is another example of something developed in one context, for one purpose,
being useful for purposes that were (probably) not foreseen.
Our intent is that this will make the technique more widely known,
and more readily used in situations where it works well.

\section*{Acknowledgments}

I thank the anonymous reviewers of an earlier version submitted to ESOP 2025 for their comments, especially for finding several technical errors.

\ifnum\OPTIONConf=2
  \bibliographystyle{abbrvnat}
\fi
\bibliography{j}

\clearpage
\appendix

\section{Bidirectional version of our example system}
\label{sec:bidir}

The type assignment system in \Figureref{fig:typing}
(let's say the system using deep subtyping;
the same issues apply with shallow subtyping)
cannot be implemented directly as a logic program:
the subsumption rule \DeepSub
\emph{always} applies,
so an implementation must use the rule only when necessary.
Mere reflexive uses of subsumption (where the type $A$ in the
typing premise is equal to the type $B$ in the conclusion) are certainly unnecessary;
moreover, since subtyping is transitive,
it is not necessary to use the subsumption rule repeatedly.
This may still leave some scope for nondeterminism,
or at least some work to develop a complete implementation strategy.
If we make the subtyping judgment more powerful, that work tends to grow.

For more powerful type systems, type assignment is often undecidable anyway.
To demonstrate that the approach considered here applies more broadly
than to this toy language,
we give a bidirectional \citep{Pierce00,Dunfield21}
system of typing rules in \Figureref{fig:typing-bi}.
Unlike the type assignment system, the bidirectional system leads directly
to a typing algorithm for sufficiently annotated terms.
Bidirectional typing splits the typing judgment $e : A$
into two: \emph{synthesis} $e => A$
and \emph{checking} $e <= A$.
Roughly, if a term $e$ has enough information to produce the type $A$,
then $e$ can synthesize $A$.
If we already know the type $e$ needs to have,
then $e$ checks against $A$.
The knowledge of the necessary type may come directly
from a type annotation, or indirectly by information flowing
through the typing derivation.

For example, the expression $\anno{(\lam{x} x)}{\Bool -> \Bool}$ synthesizes $\Bool -> \Bool$:
we check the annotated expression $\lam{x} x$ against
$\Bool -> \Bool$ by assuming that $x$ has type $\Bool$
and checking the body $x$ against $\Bool$.
Since $x$ synthesizes $\Bool$, it also checks against $\Bool$.

The design of our bidirectionalization follows the Pfenning recipe
\citep{Dunfield21} to some extent,
with some convenient departures.
The core of the recipe is that introduction rules should
check, and elimination rules should synthesize.
More specifically,
the principal judgment (the one containing the connective being introduced or eliminated)
of each introduction rule should check,
and the principal judgment of each elimination rule should synthesize.
The other judgments in each rule can be chosen to maximize typing power.

The recipe would have checking rules for the constants $\Btrue$ and $\Bfalse$,
for uniformity with other introduction forms;
we allow them to synthesize (\GenBoolIntro),
which this is convenient for programming and adds minimal
complexity.
Dually, the recipe would require the condition $e$
in $\ite{e}{e_1}{e_2}$, being the principal judgment
in an elimination rule, to synthesize.
But we make it a checking rule, providing some additional convenience.

\subsection{Bidirectional rules in detail}
Rule \GenVar synthesizes the type of $x$ by looking up $x$ in $\Gamma$.

Given a type $A -> B$, rule \GenArrIntro assumes that $x$ has type $A$
and checks the body $e$ against $B$.
A synthesis version of \GenArrIntro would be nontrivial because, even if $e$ could
plausibly synthesize the result type $B$, we would not know what to use as the argument type $A$.
That is, such a rule would not be \emph{mode-correct} \citep{Warren77,Dunfield21}.

Rule \GenArrElim requires that the function $e_1$ synthesize a type,
which provides the expected argument type $A$.
A version of this rule with a checking conclusion would be possible:
if we are checking $e_1 e_2$ against a known type $B$,
and the argument $e_2$ synthesizes its type $A$,
we have enough information to check $e_1$ against $A -> B$.

The subsumption rule \DeepSub uses the deep subtyping relation $A :< B$ to mediate between
checking and synthesis: to check that $e$ has a given type $B$,
we try to synthesize a type $A$, then use subtyping to verify that every $A$ is also a $B$.
(Even if we had no subtyping in the conventional sense,
we would need this rule, with $A = B$ in place of the subtyping premise.
If we think of subtyping as ``that which is checked by bidirectional subsumption'',
then a system without subtyping has most trivial form of subtyping,
where $A$ is a subtype of $B$ if and only if $A = B$.)

Rule \GenAnno says that to synthesize $A$ from $\anno{e}{A}$,
we check $e$ against $A$.
If we erase annotations, \GenAnno is the inverse of subsumption with trivial subtyping:
the premise of \GenAnno checks and the conclusion synthesizes;
in subsumption, the premise synthesizes and the conclusion checks.

The rules for products follow the recipe.
\GenProdIntro is an introduction rule, so we make the conclusion checking,
which provides the types needed to check the components.
\GenProdElim is an elimination rule,
so the principal judgment with $A_1 * A_2$ synthesizes,
which provides the type $A_k$ to synthesize in the conclusion.

We depart from the recipe in \GenIntIntro (really two rules, one for addition and one for subtraction):
it's an introduction rule, but for convenience, we make it synthesize.
In \GenInt we treat addition and subtraction as the application of built-in functions:
implicitly, ${+}$ synthesizes $\Int * \Int -> \Int$,
providing the type $\Int * \Int$ to check $\pair{e_1}{e_2}$ against,
which in turn requires checking $e_1$ against $\Int$ and $e_2$ against $\Int$.
(%
Similar reasoning holds in curried style, where ${+} => (\Int -> \Int -> \Int)$.)

Similarly, \GenRat synthesizes its conclusions.

Rule \GenBoolIntro synthesizes for the same reason that \GenIntIntro does.
Again departing from the recipe, \GenBoolElim checks the condition $e$ against $\Bool$,
despite being an elimination form.

\begin{figure}
  
  \judgbox{\arrayenvbl{
      \Gamma |- e => A
      \\
      \Gamma |- e <= B
      }}{
      Under typing assumptions $\Gamma$, expression $e$ synthesizes type $A$
      \\[0.45ex]
      Under typing assumptions $\Gamma$, expression $e$ checks against type $B$
      \\[0.6ex] ~
    }
    \vspace*{-2ex}
  \begin{mathpar}
    \Infer{\GenVar}
      {(x : A) \in \Gamma}
      {\Gamma |- x => A}
    \and
    \Infer{\GenArrIntro}
      {
        \Gamma, x : A |-  e <= B
      }
      {\Gamma |- \lam{x} e <= A -> B}
    ~~
    \Infer{\GenArrElim}
      {
        \Gamma |- e_1 => A -> B
        \\
        \Gamma |- e_2 <= A
      }
      {\Gamma |- e_1 e_2 => B}
  \vspace*{-0.6ex}
  \end{mathpar}

  \begin{minipage}[t]{0.43\linewidth}
   ~\\[-16ex]
  \begin{mathpar}
    \Infer{\DeepSub}
        {
          \Gamma |- e => A
          \\
          A :< B
        }
        {\Gamma |- e <= B}
  \end{mathpar}
  \end{minipage}
  \!\!\!\!\!
 \begin{minipage}[t]{0.55\linewidth}
  \begin{tcolorbox}[colback=blue!1,
                  colframe=dLightBlue,
                  width=\linewidth,
                  arc=1mm, auto outer arc,
                 ]
  \vspace*{-0.6ex}
   $~$\hspace*{-3.5ex}
      \judgbox{\arrayenvbl{
          \Gamma \sts e => A
          \\
          \Gamma \sts e <= B
          }}{
          Modification replacing \\ ${:<}$ with ${:<:}$ \\[3ex]
        }
  \vspace{2ex}
  Change $|-$ to $\sts$ throughout, \\ and replace \DeepSub with:
  \\[-1ex]
  \begin{mathpar}
    \Infer{\ShallowSub}
        {
          \Gamma \sts e => A
          \\
          A :<: B
        }
        {\Gamma \sts e <= B}
  \vspace*{-1ex}
  \end{mathpar}
  \end{tcolorbox}
  \end{minipage}

  \vspace*{-2.5ex}

  \begin{mathpar}
    \Infer{\GenAnno}
        {
          \Gamma |- e <= A
        }
        {
          \Gamma |- \anno{e}{A} => A
        }
   \and
    \Infer{\GenProdIntro}
        {
          \Gamma |- e_1 <= B_1
          \\
          \Gamma |- e_2 <= B_2
        }
        {\Gamma |- \pair{e_1}{e_2} <= B_1 * B_2}
    ~~
    \Infer{\GenProdElim}
        {
          \Gamma |- e => A_1 * A_2
        }
        {
          \Gamma |- \proj{k}{e} => A_k
        }
    \\
    \Infer{\GenIntIntro}
      {}
      {\Gamma |- n => \Int}
    \Infer{\GenInt}
      {
         \Gamma |- e_1 <= \Int
         \\
         \Gamma |- e_2 <= \Int
       }
       {
         \Gamma |- e_1 + e_2 => \Int
         \\
         \Gamma |- e_1 - e_2 => \Int
       }
    \\
    \Infer{\GenRat}
      {
         \Gamma |- e_1 <= \Rat
         \\
         \Gamma |- e_2 <= \Rat
       }
       {
         \Gamma |- e_1 + e_2 => \Rat
         ~~~
         \Gamma |- e_1 - e_2 => \Rat
         ~~~
         \Gamma |- e_1 < e_2 => \Bool
         ~~~
         \Gamma |- \rat{e_1}{e_2} => \Rat
       }
    \\
    \Infer{\GenBoolIntro}
      {b \in \{\Btrue, \Bfalse\} }
      {\Gamma |- b => \Bool}
    ~~~
    \Infer{\GenBoolElim}
        {
          \Gamma |- e <= \Bool
          \\
          \arrayenvbl{
          \Gamma |- e_1 <= B
          \\
          \Gamma |- e_2 <= B
          }
        }
        {\Gamma |- \ite{e}{e_1}{e_2} <= B}
\end{mathpar}

  \lesscaptionspace
    
  \caption{Bidirectional version of the Fig.\ \ref{fig:typing} systems, using deep and shallow subtyping}
  \label{fig:typing-bi}
\end{figure}

\end{document}